%Paper: hep-th/9406177
%From: Gabriele Ferretti <ferretti@fy.chalmers.se>
%Date: Mon, 27 Jun 94 17:40:48 +0200

%REFERENCE AND EQUATION  COUNTERS

\def\ifundefined#1{\expandafter\ifx\csname
#1\endcsname\relax}

\newcount\eqnumber \eqnumber=0
\def\beq{ \global\advance\eqnumber by 1 $$ }
\def\eeq{ \eqno(\the\eqnumber)$$ }
\def\label#1{\ifundefined{#1}
\expandafter\xdef\csname #1\endcsname{\the\eqnumber}
\else\message{label #1 already in use}\fi}
\def\(#1){(\csname #1\endcsname)}

\newcount\refno \refno=0
\def\[#1]{\ifundefined{#1}\advance\refno by 1
\expandafter\xdef\csname #1\endcsname{\the\refno}
\fi[\csname #1\endcsname]}
\def\refis[#1]{\item{\csname #1\endcsname.}}

\baselineskip=18pt
\magnification=1100
\def\tr{\;{\rm tr}\;}
\def\Tr{\;{\rm Tr}\;}
\def\res{\;{\rm Res}\;}
\def\sym{\;{\rm Sym}\;}
\def\gl1{{\bf\rm gl}_{\rm res}}
\def\map{{\rm Map}(M,{\bf g})}

%%%%%%%%%%%%%%%%%%%%%%%%%%%%%%%%%%%%%%%%%%%%%%
\noindent ITP 94-17

\noindent{\tt hep-th/9406177 }
\vskip.5in

\centerline{{\bf REGULARIZATION AND QUANTIZATION}}
\centerline{{\bf OF}}
\centerline{{\bf HIGHER DIMENSIONAL CURRENT ALGEBRAS }\footnote\S{Talk
            given at the G\"ursey Memorial Conference I on Strings and
            Symmetries, Istanbul, Turkey, June 6-10 1994.}}
\vskip 1in

\centerline{\bf G. Ferretti}
\vskip.5cm
\centerline{\sl Institute of Theoretical Physics}
\centerline{\sl Chalmers University of Technology}
\centerline{\sl S-41296 G\"oteborg, Sweden}
\centerline{\tt ferretti@fy.chalmers.se}
\vskip 1in
\centerline{\bf Abstract}

We present some recently discovered infinite dimensional Lie algebras that can
be understood as extensions of the algebra
$\map$ of maps from a compact $p$-dimensional manifold $M$ to some finite
dimensional Lie algebra ${\bf g}$. In the first part of the paper, we describe
the physical motivations for the study of these algebras. In the second part,
we discuss their realization in terms of pseudo-differential operators and
comment on their possible representation theory.

\vfill\eject
\centerline{{\bf Why study} $\map$ {\bf~ and its extensions?}}

Let us start with a situation that is familiar to everybody and consider the
one
dimensional case, $M=S^1$, with ${\bf g}$ some finite dimensional Lie algebra.
There, ${\rm Map}(S^1,{\bf g})$ has a unique
central extension $\widehat{{\rm Map}}(S^1,{\bf g})$ given by the cocycle
\beq
        c(\lambda,\xi) = {i\over{2\pi}}\int_0^{2\pi} \tr
             (\lambda(x)\xi^{\prime}(x) ) dx, \label{central}
\eeq
which is known as the (untwisted) affine Kac-Moody algebra [Ba-71] [Ka-67]
[Ka-85b] [Mo-67] [Mi-89] [Wi-84]. An element of the algebra can be written as a
pair $(\lambda;z)$ where $\lambda$ is a map from the circle into ${\bf g}$ and
$z$ is a complex number. The Lie product on these pairs is
\beq
        [(\lambda;z),(\xi;w)] = ( [\lambda,\xi]; c(\lambda,\xi)).
         \label{kacmoody}
\eeq

Consider now the normal ordered currents
$J^a(x)=(i/2) :\Psi^T(x) M^a \Psi(x):$ where
the $M^a$ are real antisymmetric generators of ${\bf g}$ normalized as
$\tr M^a M^b = -2k\delta^{ab}$ and $\Psi(x)$ are Majorana-Weyl fermionic fields
on the light-cone. The currents $J^a$ satisfy the celebrated current algebra
\beq
       [J^a(x),J^b(y)]= if^{abc}J^c(x)\delta(x-y) +{{ik}\over{2\pi}}
       \delta^{ab}\delta^{\prime}(x-y), \label{onedim}
\eeq
where the second term on the r.h.s. is the Schwinger term arising from normal
ordering.

Noticing that the currents in \(onedim) behave as operator valued scalar
densities of weight one, we can then take the product
$J(\lambda)=\int dx\; J^a(x)\lambda^a(x)$ to obtain a representation of the
affine algebra \(kacmoody) in the fermionic Fock space:
\beq
         [J(\lambda)+k z,J(\xi)+k w] = J([\lambda,\xi]) +
         k c(\lambda,\xi).
\eeq
Therefore, the study of representations of the algebra \(kacmoody) is
equivalent to the study of $(1+1)$-dimensional current algebras \(onedim).
It is then clear that
current algebras in higher dimensions must be related to some extensions
of the algebra $\map$.

\centerline{{\bf What are the ``interesting'' extensions of}
                    $\map${\bf~ when~} $p > 1${\bf ?}}

When considering higher dimensional manifolds we first encounter the fact that,
contrary to the one dimensional case, $\map$ has an infinite number of central
extensions [Pr-86]. Roughly speaking, for any loop one can draw on the manifold
(even contractable loops) and for any two elements $\lambda,\xi \in \map$,
there is a two cocycle analogous to \(central) given by evaluating the
pull-back of the form $\tr(\lambda d\xi)$ on such loop. These are essentially
all the possible central extensions one can have. However, none of these
extensions are interesting in relation to the study of higher dimensional
current algebras precisely because of their ``one dimensional'' nature.

We must therefore look for more general kinds of extensions (abelian or even
non abelian) and this always means adding extra
fields to the system. Also, if we are aiming at an algebra that does not depend
on the detailed geometry of the manifold $M$, (e.g. its metric), it is a
natural guess to consider adding differential forms. In
order to keep the discussion contained, we will now present the algebra in an
axiomatic way and, after that, argue about its relevance to physical systems of
interest.

\centerline{{\bf Our proposal of extension of} $\map${\bf .}}

The algebra we are proposing is made of pairs $(X,z)$, where $X$ represents the
free sum
$X=X^{(0)} + (X^{(2)} + X^{(4)} + \cdots X^{([p])})$, $X^{(0)}\in \map$ and
$X^{(2)}, X^{(4)},\cdots $ are ${\bf g}$ valued differential forms on $M$ of
even degree
$2,4,\cdots$. The symbol $[p]$ denotes the largest even integer less or equal
than $p=\dim (M)$ and $z$ is a complex number.
In the following, ${\bf g}$ is
taken to be either ${\bf gl}(n)$ or ${\bf u}(n)$.

The Lie product in such an algebra can be written concisely as
\beq
        [(X;z) , (Y;w)] = ([X,Y] + [dX,dY] ; {i\over{(2\pi)^p}}\int_M
            \tr(XdY)), \label{ouralgebra}
\eeq
where the following conventions have been made:
\item{a.} The wedge product is understood in all the commutators of forms on
the r.h.s.
\item{b.} Forms of degree larger that $p$, arising from the wedge
products, all vanish.
\item{c.} The integral on the r.h.s. vanishes on forms of degree not equal to
$p$. (In particular it always vanishes when $p$ is even.)

\noindent We will now argue that algebras of this kind are of interest in
higher dimensional physics by studying a few examples in various dimensions.
\item{\bf $p=1$} In this case the algebra \(ouralgebra) obviously reduces to
the affine Kac-Moody algebra \(kacmoody) presented in the introduction.
\item{\bf $p=2$} This is the current algebra of planar QCD in the parity
conserving phase [Fe-92] [Fu-94] (i.e., with an even number of flavors $N_f$
and a large number of colors $N_c$). Explicitly, in terms of the flavor
currents $J^a(x)$ and the Goldstone field $\Phi^a(x)$, it takes the form (up to
some rescaling)
\beq\eqalign{
        [J^a(x), J^b(y)] &= i f^{abc}J^c(x)\delta(x-y) +
         {{i N_c}\over{4\pi}}d^{abc}\epsilon^{\mu\nu}\partial_\mu
                                   \Phi^c(x)\partial_\nu\delta(x-y)\cr
        [J^a(x), \Phi^b(y)] &= i f^{abc} \Phi^c(x)\delta(x-y)\cr
        [\Phi^a(x), \Phi^b(y)]  &=0.\cr}\label{QCD2}
\eeq
\item{\bf $p=3$} This is probably the most interesting case, because, in three
dimensions, \(ouralgebra) describes the gauge algebras arising from the
canonical formulations of anomalous gauge theories with chiral fermions
[Mi-83] [Fa-84a] [Fa-84b] [Mi-85] [Pe-88] [Fl-89] [Jo-85]. These algebras are
known in the literature
under the name of Mickelsson-Faddeev-Shatasvili (MFS) algebras. In their most
explicit form, as linear algebras in the gauge generators
$G^a(x)$ and the gauge fields $A^a_\mu(x)$ they can be written as [Fa-84a]:
\beq\eqalign{
        [G^a(x), G^b(y)] &= i f^{abc}G^c(x)\delta(x-y) +
         {1\over{12i\pi^2}}d^{abc}\epsilon^{\mu\nu\rho}\partial_\mu
            A^c_\nu (x)\partial_\rho\delta(x-y)\cr
        [G^a(x), A^b_\mu (y)] &= i f^{abc} A^c_\mu (x)\delta(x-y)
            +\delta^{ab}\partial_\mu\delta(x-y)\cr
        [A^a_\mu(x), A^b_\nu(y)]  &=0.\cr}\label{MFS}
\eeq
The gauge field $A^a_\mu$ is associated with the two form $X^{(2)}$ and the
scalar density $G^a$ with the scalar  $X^{(0)}$. There is also a connection
with the BRST algebra of three dimensional extended objects ($3$-branes)
[Di-92]
\footnote\S{In general, these algebras do not
arise as Noether symmetries of the $p$-brane
action [Pe-93] but they can still be used to construct a BRST operator for
the $p$-brane functional.}.
\item{\bf $p=4$} This is the lowest dimension in which \(ouralgebra)
becomes a non-abelian extension of $\map$ (the wedge product of two 2-forms is
a 4-form). More interesting is the case:
\item{\bf $p=5$} This algebra has been studied in connection with the bosonic
sector of the $5$-brane [Ce-94], believed to be of interest in
the study of non-perturbative effects in string theory [Se-94].
It is an alternative and inequivalent formulation of the problem studied in
[Di-93]. It should also be stressed that for odd dimensions higher than three,
the algebras obtained from \(ouralgebra) are not equivalent to the higher
dimensional versions of the MFS algebras nor can the two algebras be related by
contraction. In this particular case, \(ouralgebra) yields:
\beq\eqalign{
[T^a(x),T^b(y)]&= if^{abc}T^c(x)\delta(x-y)+
	d^{abc}\partial_i T^{c\,ij}(x)\partial_j\delta(x-y)\cr
[T^a(x),T^{b\,mn}(y)]&= if^{abc}T^{c\,mn}(x)\delta(x-y)-
	d^{abc}\epsilon^{ijkmn}\partial_iT^c_j(x)\partial_k
        \delta(x-y)\cr
[T^a(x),T^b_i(y)]&= if^{abc}T^c_i(x)\delta(x-y)+
	k\delta^{ab}\partial_i\delta(x-y)\cr
[T^{a\,kl}(x),T^{b\,mn}(y)]&= if^{abc}\epsilon^{iklmn}T^c_i(x)
         \delta(x-y)+ k\delta^{ab}\epsilon^{iklmn}\partial_i
         \delta(x-y)\cr
[T^{a\,mn}(x),T^b_i(y)]&=0\cr
[T^a_i(x),T^b_j(y)]&=0,\cr}\label{pbrane}
\eeq
in terms of a set of generators $T^a(x)$, $T^{a\,ij}(x)$, $T^a_i(x)$ and $k$.
(See [Ce-94] for details.)

\centerline{\bf Regularization of the Algebra}

Having presented the algebra, we will now examine its explicit realization as
algebra of operators in some Fock space. For definitiveness, let us focus on
the three dimensional case, where one can write the Lie product of
$\widehat\map$ as
\beq\eqalign{
        [(X^{(0)} + X^{(2)};z) , (Y^{(0)} + Y^{(2)};w)]=&([X^{(0)},Y^{(0)}] +
        [X^{(0)},Y^{(2)}] + [X^{(2)},Y^{(0)}] + \cr &[dX^{(0)},dY^{(0)}];
        {i\over{(2\pi)^3}} \int_M \tr (X^{(0)}dY^{(2)} + X^{(2)}dY^{(0)})).\cr}
         \label{ourthree}
\eeq
The simplest and most obvious choice for the Fock space is a fermionic Fock
space $\cal F$ constructed by considering Weyl spinor fields $\Psi(x)$ at equal
time $t=0$. (Contrary to $1+1$ dimensions, there are no Majorana-Weyl spinors
in $3+1$ dimensional space time.) Using these fields one can easily construct
the charge densities (i.e. time components of the currents):
\beq
       J^a(x) = {i\over 2} : \Psi^\dagger(x) M^a \Psi(x): \;\;\; ,
        \label{chargedensity}
\eeq
Where $M^a$ are now antihermitian generators of ${\bf g}$. If, in analogy with
the one dimensional case, we tried to set $J(X^{(0)})=
\int_M J^a(x) X^{(0) a}(x) d^3x$, we would immediately run into the following
problem: Contrary to the one dimensional case,
the operator $J(X^{(0)})$ creates states of infinite norm out of the ordinary
vacuum \footnote\S{Even in one dimension, there are cases where normal
ordering is not enough [Ev-94] but they do not arise in this context.}.
A few years ago, Mickelsson and Rajeev proposed to enlarge the Fock space to
``make room'' for these extra states [Mi-88] [Fu-90] [La-91] [La-94].
Unfortunately, it was later realized by Pickerell that there is a no-go theorem
that prevents
one from finding unitary representations this way [Pi-87] [Pi-89].

More recently, Mickelsson has proposed a different approach
[Mi-93a] [Mi-93b] [Mi-94]. The basic idea is to regularize the argument of $J$
in a way that, on one hand, preserves the algebra and, on the other hand, makes
the norms converge. Originally the idea was applied to the MFS algebra with
explicit dependence of the regulators on the gauge fields. This creates some
problems in the implementation, and one has to content oneself with finding a
projective representation.
We follow a simpler approach, where
all generators are regularized by the action of a linear operator and there is
no explicit dependence on the gauge fields.

Let then $A : {\cal H} \to {\cal H} $ be an arbitrary operator in the first
quantized Hilbert space of Weyl spinorial wave functions $\psi\in{\cal H}$.
(E.g. when $A=X^{(0)}$ we think of it as a multiplicative operator.) Also, let
$h$ be the helicity operator, which coincides with
$(-1)\times$ the sign of the energy. It is a well known calculation that shows
that
\beq
       || J(A) |0>||^2 = \Tr_{{\cal H}}( [h,A]^\dagger[h,A]).
       \label{normtrace}
\eeq
This calculation follows straightforwardly by writing the l.h.s. in terms of
the fermionic oscillators and expanding the normal ordered
bilinears. Expressed in more mathematical language, what \(normtrace) says is
that $J(A) |0>$ has finite norm iff $[h,A]$ is an Hilbert-Schmidt operator,
i.e. the trace of its square is finite.

To be more explicit, let $A$ be a pseudo-differential operator (PSDO)
[Co-85] [Co-88] [Gu-85] [La-89] [Va-92] defined through its {\it symbol}
$a(x,p)$, function on the phase space:
\beq
        \sym(A) = a(x,p) \approx \sum_{k=0}^\infty a_{-k}(x,p) \quad
        \hbox{where}\quad a_{-k}(x,p) \approx {1\over{|p|^k}}, \quad
        a_{-k}=O(-k).
\eeq
(We do not allow for positive powers of $p$ in the asymptotic expansion of the
symbol.)
The relation between the operator $A$ and its symbol $a$
is given by defining the action of  $A$ on ${\cal H}$ as:
\beq
        A\psi(x) = {1\over{(2\pi)^{3/2}}} \int e^{ip\cdot x} a(x,p)
                          \tilde\psi(p) d^3p,
\eeq
where $\tilde\psi$ is the ordinary Fourier transform of $\psi$.
For example, the symbol of $X^{(0)}$ as  a multiplicative operator is
$X^{(0)}$ itself as a function of $x$, whereas the symbol of $h$ is
$\sym(h)\equiv\epsilon=(p_\mu\sigma_\mu)/|p|$, where $\sigma_\mu$ are the
ordinary Pauli matrices. The algebra of
PSDO's of this kind is defined through the {\it star product} $*$
\beq
       a*b = \sym(AB) = a(x,p) \exp(-i {{\leftarrow\partial}\over
       {\partial p_\mu}}\cdot{{\partial\rightarrow}\over
       {\partial x^\mu}}) b(x,p).
\eeq
One can then restate the Hilbert-Schmidt condition in terms of the
ultra-violet asymptotic behavior of the symbols and say that $J(A)|0>$ has
finite norm iff $[\epsilon,a]_* \approx p^{-2} \equiv O(-2)$ (square integrable
in $p\in {\bf R}^3$.) We shall denote the space of these operators as
${\cal W}$.

The multiplicative operator $X^{(0)}$ does not have this property, being the
$*$-commutator with $\epsilon$ only of degree $O(-1)$. We must therefore add a
counterterm of degree $O(-1)$ and try to cancel the offending term.
Define:
\beq
      \theta(X^{(0)}) = X^{(0)} +
      S_\mu{{\partial X^{(0)}}\over{\partial x^\mu}} + O(-2).
\eeq
then, by using the asymptotic behavior of the symbols,
\beq
     [\epsilon, \theta(X^{(0)})]_* = O(-2) \quad\hbox{iff}\quad
     [\epsilon, S_\mu {{\partial X^{(0)}}\over{\partial x^\mu}}]=
     i{{\partial\epsilon} \over{\partial p_\mu}}
     {{\partial X^{(0)}}\over{\partial x^\mu}}, \label{firstdegree}
\eeq
where the commutator in the r.h.s. of \(firstdegree) is an ordinary matrix
commutator.
Equation \(firstdegree) has a solution that can be written as, up to
unimportant terms that commute with $\epsilon$ [Mi-93a] [Mi-94],
\beq
     S_\mu={{\epsilon_{\mu\nu\rho} p_\nu \sigma_\rho}\over{2p^2}}
     \label{smu}
\eeq
where $\epsilon_{\mu\nu\rho}$ denotes the totally antisymmetric tensor.
This choice makes $\theta(X^{(0)})$ into an operator with a good second
quantized picture (i.e. an element of ${\cal W}$) and preserves the original
algebra to degree $O(-1)$.
At this point one may wonder how far one should go and how to incorporate the
term  $X^{(2)}$ into the picture. We answer both questions by considering
the central term of the algebra.

\centerline{\bf The twisted Radul cocycle}

In the space of PSDO's there is essentially only one trace we can define. This
is given by the so called Wodzicki residue [Wo-85], a higher dimensional
version of the Adler-Manin residue [Ad-79] [Ma-79] that, on a three dimensional
manifold, can be written as
\beq
    \res(a) = {1\over{(2\pi)^3}} \int_{|p|=\Lambda}\tr a_{-3}(x,p)
             \eta \wedge d\eta \wedge d\eta, \label{wodzicki}
\eeq
where $\eta=p_\mu dx^\mu$ is the canonical one-form on the cotangent bundle
of $M$. We assume that the symbol has compact support in $x$ to avoid
considering global properties. Using this trace, one can write cochains in the
space of PSDO's. In particular, we need a two cocycle that is a version of the
so called Radul cocycle [Ra-91a] [Ra-91b] twisted
by the presence of a factor $\epsilon$ [Mi-94]. The Radul cocycle is, again, a
higher dimensional generalization of the one-dimensional Kravchenko-Khesin
cocycle [Kr-91] that we define as follows:
\beq
     c_R(a,b) = {{3i}\over\pi}\res(\epsilon*[\log|p|,a]_* *b). \label{radul}
\eeq
Looking at \(wodzicki) we see that the trace depends on the terms of degree
$O(-3)$ in the asymptotic expansion. However, because of the presence of
$\log|p|$ in \(radul), it is easy to see that the {\it cohomology} only depends
on terms up to degree $O(-2)$. The result of this analysis is that, in order to
have a non trivial central term, which we expect to be generated at the quantum
level by rearranging the normal ordered fermions, one has to go one more step
and compute the terms of degree $O(-2)$.

This turns out to be important also for another reason. In order to have the
regularized operators satisfying the original algebra \(ourthree) to degree
$O(-2)$, one {\it must} include a two-form
realized as a PSDO of degree $O(-2)$; it is simply impossible to obtain a
regulator that works to this degree by using the scalars $X^{(0)}$ alone.
This is very pleasing because it gives a reason for the forms of higher degree
to be included in the algebra. Finally, we would like to point out that this
way of proceeding breaks down for space dimensions strictly higher than three
(i.e. quantum fields in space-time dimensions strictly higher
than $3+1$), indicating some sort of ``upper critical dimension'' for this
regularization procedure. Without giving all the details of the calculations
(to be presented in a forthcoming paper), we shall present the final answer for
the regulated algebra:
\beq
    \theta(X^{(0)}+X^{(2)})=X^{(0)}+S_\mu{{\partial X^{(0)}}
    \over{\partial x^\mu}} +
    {1\over 2} S_{\mu\nu} {{\partial}\over{\partial x^\mu}}
                          {{\partial X^{(0)}}\over{\partial x^\nu}}+
    {1\over 2} A_{\mu\nu} X^{(2)}_{\mu\nu}, \label{final}
\eeq
where $S_\mu$ was given in \(smu) and the other two terms can be expressed as
functions of $S_\mu$ as
\beq\eqalign{
     S_{\mu\nu} &= {1\over 2}(S_\mu S_\nu + S_\nu S_\mu) -{i\over 2}
     ({{\partial S_\mu}\over{\partial p^\nu}} +
      {{\partial S_\nu}\over{\partial p^\mu}})\cr
     A_{\mu\nu} &= {1\over 2}(S_\mu S_\nu - S_\nu S_\mu) -{i\over 2}
     ({{\partial S_\mu}\over{\partial p^\nu}} -
      {{\partial S_\nu}\over{\partial p^\mu}}).\cr}\label{degreetwo}
\eeq
It can then be checked by an explicit calculation that eq. \(final) satisfies
all the requirements for our regulated algebra. In particular, not only is the
$*$-commutator of $\theta(X^{(0)}+X^{(2)})$ with $\epsilon$ defining an
Hilbert-Schmidt operator, but also the Lie product \(ourthree) is preserved to
degree $O(-2)$
\beq
      [\theta(X^{(0)}+X^{(2)}),\theta(Y^{(0)}+Y^{(2)})]_*=
       \theta([X^{(0)},Y^{(0)}] + [X^{(2)},Y^{(0)}] + [X^{(0)},Y^{(2)}] +
       [dX^{(0)},dY^{(0)}]), \label{liereg}
\eeq
and the twisted Radul cocycle \(radul) generates the correct central term
after integrating out the momentum variables
\beq
    c_R(\theta(X^{(0)}+X^{(2)}),\theta(Y^{(0)}+Y^{(2)})) =
    {i\over{(2\pi)^3}} \int_M \tr (X^{(0)}dY^{(2)} + X^{(2)}dY^{(0)}).
\eeq

Since the $*$-product of two PSDO's always generates terms of arbitrary low
degree, if one wants to interpret $\theta$ as a
Lie algebra homomorphism, one has
to quotient out an ideal from the space of PSDO's
under consideration. In other
words, the homomorphism $\theta$ is defined only on the equivalence classes of
operators that differ by an (uninteresting) PSDO of degree $O(-3)$ or less. We
denote this homomorphism by
\beq
    \theta : \widehat\map \rightarrow \widehat{\cal W}/{\cal I}_{-3},
\eeq
where $\widehat{\cal W}$ is the extension, via the cocycle \(radul), of the
algebra ${\cal W}$ defined above and
${\cal I}_{-3}$ is the ideal of ${\cal W}$ consisting of PSDO's of degree
$O(-3)$ or less.
Notice that the quotient $\widehat{\cal W}/{\cal I}_{-3}$ is well defined
because the cocycle \(radul)
vanishes if one of its arguments is in the ideal ${\cal I}_{-3}$.

\centerline{\bf Conclusions and further projects}

In conclusion, we have presented some extensions of the algebra $\map$ that
have
applications in various fields of physics (current algebras, gauge anomalies,
$p$-branes etc...). For the three dimensional case, we have found a way of
regularizing these algebras that allows one to have well defined second
quantized operators in a fermionic Fock space, and ``looks promising'' as far
as the representation theory is concerned.

Obviously, much work still needs to be done. We would like to emphasize two
points that deserve further investigations: The first is the need to gain a
better understanding of how the central term can arise from second
quantization. We know that the cocycle
\(radul) must somehow be related to the universal cocycle of ${\bf gl}_1$
[Ka-85a] [Pr-86]
[Lu-76] but we would
like to have an explicit derivation in more
physical terms. The second point is
to better understand the representation theory of these algebras [Ki-76], in
particular,
to understand how, and if, it is possible to quotient the representations of
$\widehat{\cal W}$ by its ideal ${\cal I}_{-3}$ and
how to use these results in
the study of some specific quantum field theories.

\centerline{\bf Acknowledgments}

This work has been done collaboration with the following people:
M. Cederwall, B.E.W. Nilsson, A. Westerberg
and J. Mickelsson. A more detailed
version of this work, addressing the issues mentioned in the concluding
remarks, is under preparation.

I would like to thank the organizers of this conference for giving me the
opportunity of this talk and for their kind hospitality during my stay in
Istanbul.

\vfill\eject

\centerline{\bf References}

\item{[Ad-79]} M. Adler, Invent. Math. 50 (1979) 219.
\item{[Ba-71]} K. Bardakci and M.B. Halpern, Phys. Rev. D3 (1971) 2493.
\item{[Ce-94]} M. Cederwall, G. Ferretti, B.E.W. Nilsson and A. Westerberg,
               Nucl. Phys. B, to appear. {\tt hep-th/9401027}
\item{[Co-85]} A. Connes, Publ. Math. IHES 63 (1985) 257.
\item{[Co-88]} A. Connes, Comm. Math. Phys. 117 (1988) 117.
\item{[Di-92]} J.A. Dixon and M.J. Duff, Phys. Lett. B296 (1992) 28.
\item{[Di-93]} J.A. Dixon, Nucl. Phys. B407 (1993) 73. {\tt hep-th/9206082}
\item{[Ev-94]} M. Evans, Talk delivered at this conference.
\item{[Fa-84a]} L. Faddeev, Phys. Lett. B145 (1984) 81.
\item{[Fa-84b]} L. Faddeev and S. Shatasvili, Theor. Math. Phys. 60 (1984) 770.
\item{[Fe-92]} G.Ferretti and S.G. Rajeev, Phys. Rev. Lett. 69
               (1992) 2033. {\tt hep-th/9207039}
\item{[Fl-89]} R. Floreanini and R. Percacci, Int. J. Mod. Phys. 17
               (1989) 4581.
\item{[Fu-90]} K. Fujii and M. Tanaka, Comm. Math. Phys. 129 (1990) 267.
\item{[Fu-94]} K. Fujii, Comm. Math. Phys. 162 (1994) 273.
\item{[Ge-83]} E. Getzler, Comm. Math. Phys. 92 (1983) 163.
\item{[Gu-85]} V. Guillemin, Adv. Math. 55 (1985) 131.
\item{[Ka-67]} V.G. Kac, Funct. Anal. App. 1 (1967) 328
\item{[Ka-85a]} V.G. Kac and D.H. Peterson, in ``Proceedings of the Summer
                School on Complete Integrable Systems'', Montreal, Canada
                 (1985).
\item{[Ka-85b]} V.G. Kac {\it Infinite Dimensional Lie Algebras}
               Cambridge University Press, Cambridge U.K (1985).
\item{[Ki-76]} A.A. Kirillov, {\it Elements of the Theory of Representations}
               Springer-Verlag, New York, (1976).
\item{[Kr-91]} O.S. Kravchenko and B.A. Khesin, Funct. Anal.
                 Appl. 25 (1991) 83.
\item{[Jo-85]} S.G. Jo, Phys. Lett. B163 (1985) 353.
\item{[La-89]} H. Lawson and M.L. Michelsohn, {\it Spin Geometry} Princeton
               University Press, Princeton (1989).
\item{[La-91]} E. Langmann, in ``Topological and Geometrical Methods in Field
                Theory'', Turku, Finland (1991).
\item{[La-94]} E. Langmann, Comm. Math. Phys. 162 (1994) 1.
\item{[Lu-76]} L.E. Lundberg, Comm. Math. Phys. 50 (1976) 103.
\item{[Ma-79]} Y.I. Manin, J. Sov. Math. 11 (1979) 1.
\item{[Mi-83]} J. Mickelsson, Lett. Math. Phys. 7 (1983) 45.
\item{[Mi-85]} J. Mickelsson, Comm. Math. Phys. 97 (1985) 361.
\item{[Mi-88]} J. Mickelsson and S.G. Rajeev, Comm. Math. Phys. 116
               (1988) 365.
\item{[Mi-89]} J. Mickelsson, {\it Current Algebras and Groups},
               Plenum, New York, (1989).
\item{[Mi-93a]} J. Mickelsson, in ``Proceedings of the XXII Int. Conf. on
               Differential Geometric Methods in Theoretical Physics'',
              Ixtapa, Mexico (1993). {\tt hep-th/9311170}
\item{[Mi-93b]} J. Mickelsson, Lett. Math. Phys. 28 (1993) 97.
              {\tt hep-th/9210069}
\item{[Mi-94]} J. Mickelsson, Stockholm Royal Inst. Tech. Preprint (1994).
              {\tt hep-th/9404093}
\item{[Mo-67]} R.V. Moody, Bull. Amer. Math. Soc. 73 (1967) 217.
\item{[Pe-88]} R. Percacci and R. Rajaraman, Phys. Lett. B201 (1988) 256.
\item{[Pe-93]} R. Percacci and E. Sezgin, Int. J. Mod. Phys. A8 (1993) 5367.
               {\tt hep-th/9210061}
\item{[Pi-87]} D. Pickrell, J. Funct. Anal. 70 (1987) 323.
\item{[Pi-89]} D. Pickrell, Comm. Math. Phys., 123 (1989) 617.
\item{[Pr-86]} A. Pressley and G. Segal, {\it Loop Groups} Oxford Science
                Publications, Oxford, (1986).
\item{[Ra-91a]} A.O. Radul, Phys. Lett. B265 (1991) 86.
\item{[Ra-91b]} A.O. Radul, Funct. Anal. Appl. 25 (1991) 25.
\item{[Se-94]} E. Sezgin, Talk delivered at this conference.
\item{[Va-92]} J.C. V\'arilly and J.M. Garcia-Bond\'ia Preprint
               DFTUZ-92-20 (1992).
\item{[Wi-84]} E. Witten, Comm. Math. Phys. 92 (1984) 455.
\item{[Wo-85]} M. Wodzicki, in ``Lecture Notes in Mathematics'', Berlin,
               Germany (1985).

\bye